\documentclass[aip,amsmath,amssymb,reprint]{revtex4-1}

\usepackage{graphicx}
\usepackage{dcolumn}
\usepackage{bm}
\usepackage{textcomp}

\usepackage[utf8]{inputenc}
\usepackage[T1]{fontenc}
\usepackage{mathptmx}
\graphicspath{ {./Figures/} }
\usepackage{ulem}

\draft 

\begin{document}

\title{Low-loss, high-bandwidth fiber-to-chip coupling using capped adiabatic tapered fibers} 

\author{Saeed Khan}
\email[]{saeed.khan@nist.gov}
\author{Sonia M. Buckley}
\author{Jeff Chiles}
\author{Richard P. Mirin}
\author{Sae Woo Nam}
\author{Jeffrey M. Shainline}

\affiliation{National Institute of Standards and Technology, 325 Broadway, Boulder, Colorado 80305, USA \looseness = -1}

\date{\today}

\begin{abstract}
We demonstrate adiabatically tapered fibers terminating in sub-micron tips that are clad with a higher-index material for coupling to an on-chip waveguide. This cladding enables coupling to a high-index waveguide without losing light to the buried oxide. A technique to clad the tip of the tapered fiber with a higher-index polymer is introduced. Conventional tapered waveguides and forked tapered waveguide structures are investigated for coupling from the clad fiber to the on-chip waveguide. We find the forked waveguide facilitates alignment and packaging, while the conventional taper leads to higher bandwidth. The insertion loss from a fiber through a forked coupler to a sub-micron silicon nitride waveguide is 1.1\,dB and the 3\,dB-bandwidth is 90\,nm. The coupling loss in the packaged device is 1.3 dB. With a fiber coupled to a conventional tapered waveguide, the loss is 1.4\,dB with a 3\,dB bandwidth extending beyond the range of the measurement apparatus, estimated to exceed 250\,nm. 
\end{abstract}


\maketitle 

\section{\label{sec:Introduction} Introduction}
A major challenge in the field of integrated photonics is the efficient coupling of light from a standard optical fiber to on-chip photonic waveguides and devices. For applications in digital computing and communications, these losses lead to significant contributions to the link power budget. For applications in quantum optics and quantum communication, losses on the order of a few percent may eliminate any quantum advantage. The problem of fiber-to-chip coupling is difficult because the fundamental mode of an optical fiber is roughly 10\,\textmu m in diameter, and the dimensions of the fundamental mode of a high-index-contrast waveguide are often less than 1\,\textmu m across. 

Several techniques have been utilized to address the issue of fiber-to-chip coupling \cite{Rev1,Rev2,Rev3,Rev4}, and different approaches may be more appropriate depending on the application. The most common techniques include grating couplers \cite{Grating1,Grating2,Grating3}, end-coupling \cite{End1,End2}, and adiabatic tapers \cite{Adiabatic1,Adiabatic2,rasc1998,Adiabatic4}. Grating couplers have the advantage of a compact footprint, but they are sensitive to alignment, and they utilize a fiber connecting to the chip at near-normal incidence. Packages become bulky out of the plane of the chip. The use of angle-polished facets mitigate this packaging problem to some extent \cite{Grating4}. Additionally, achieving fiber-to-chip insertion loss of a few percent with grating couplers is difficult, even in theory \cite{miya2018}. The highest-efficiency grating couplers utilize two independently patterned layers, with state-of-the-art planarization between them \cite{Grating3}. Such structures are not easily realizable in all integrated photonics processes.  End couplers differ from grating couplers in that the direction of light propagation is not required to change, leading to low loss with high bandwidth in a simple geometry. Additionally, the fibers are aligned in the plane of the chip, resulting in smaller package size. The downside of end coupling is that mode matching is required at an interface, often requiring precise alignment between the fiber and the on-chip waveguide. In one variant on this approach, an on-chip, index-engineered waveguide converts between the large mode in the fiber and the compact mode on the chip\cite{End1}. To maintain the mechanical integrity of the packaged assembly, V-grooves are etched in the chip leading up to the waveguide. These V-grooves are usually etched with hot potassium hydroxide and extend into the surface of the silicon chip by roughly the fiber radius (62.5\,\textmu m). Again, this fabrication is not compatible with all integrated photonics processes.

In this work, we explore a variation on the concept of terminating tapered fibers to investigate whether they may enable low loss, high bandwidth, and straightforward packaging compatible with a variety of integrated photonics processes. We differentiate between terminating tapered fibers (also referred to as conical tapers), which end with a fine tip, and continuous tapered fibers, wherein a single-mode fiber is drawn to a narrow but continuous section that exposes the evanescent field \cite{bili1992,knch1997}. While continuous tapers are useful for coupling into a variety of photonic cavity structures \cite{srba2004,srpa2007}, we focus here on terminating tapered fibers for coupling to a single-mode waveguide located on a buried oxide (BOx) under-cladding. The objective is to realize low loss and high optical bandwidth couplers in a high-throughput packaging technology compatible with manufacturing into fiber ribbons.

Adiabatic couplers are conducive to high-bandwidth, low-loss operation. In adiabatic fiber-to-chip couplers, both the optical fiber and the on-chip waveguide are slowly tapered along the propagation direction while the optical power remains in the fundamental mode of the fiber-waveguide structure. Because neither mode matching nor $k$-vector matching at an abrupt interface occurs, loss can be low across a broad bandwidth without high sensitivity to position.  With a similar approach, single-mode fiber-waveguide coupling efficiencies as high as 97\,\% (-0.13\,dB) have been demonstrated \cite{Adiabatic1}, although in that work the waveguides were suspended in air. The requirement that the waveguide be suspended limits the domain of applicability of this method of fiber coupling. Here we report an approach to overcome this limitation by cladding the tapered fiber with a high-index polymer. In this coupling scheme, light is adiabatically transferred from the fundamental mode of an optical fiber, through a tapered region, into a higher-index cladding formed by SU8 polymer, and finally into a sub-micron SiN waveguide. By covering the fiber tip in a material with higher index of refraction than SiO$_2$ we avoid losing light to the continuum of modes in the BOx. We consider both tapered and forked couplers in the on-chip waveguide, and we describe the technique to clad the tip of the tapered fiber with the polymer. We also present a method for packaging the tapered fibers with the integrated photonic chip. By using a separate carrier chip for the fiber, the method decouples the processing of grooves that hold the fiber from the fabrication of the photonic circuits. 
    
\section{\label{sec:Principle} Principle of Operation}
\begin{figure}
\includegraphics[width=\linewidth]{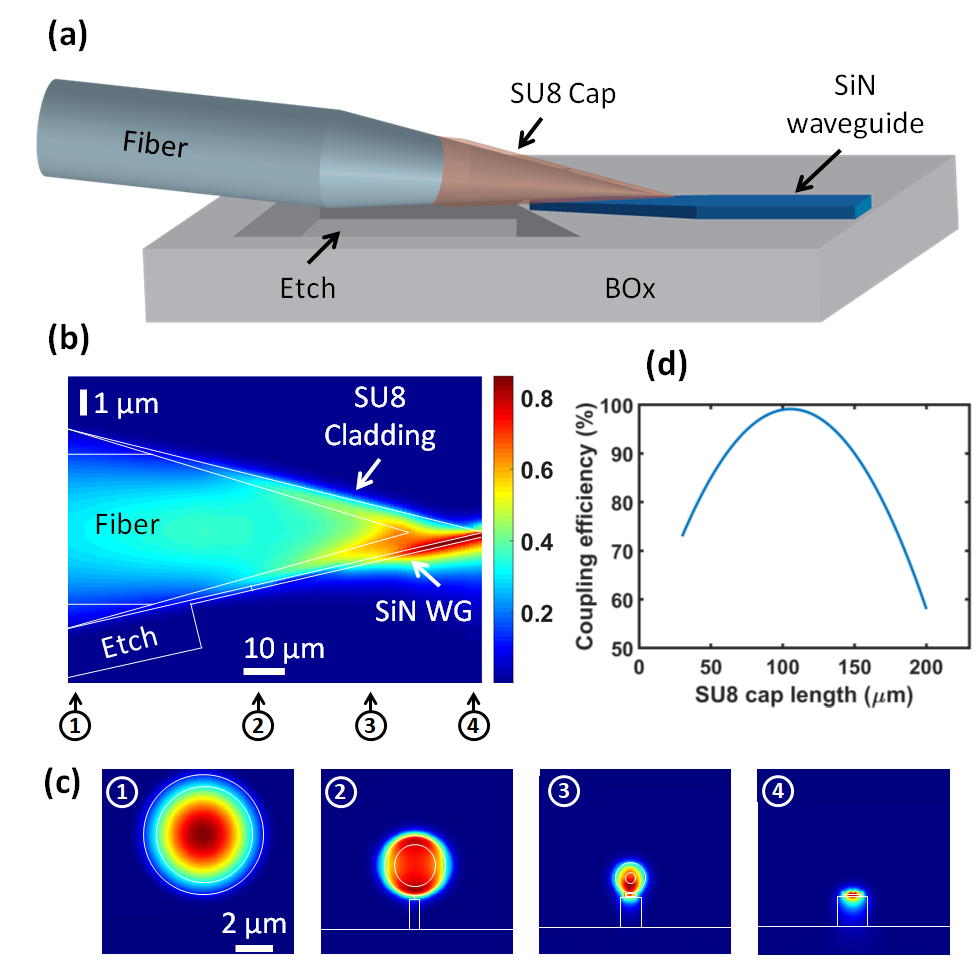}
\caption{\label{fig:1stSchematic} (a) Schematic of adiabatic coupling from tapered optical fiber to waveguide (WG).  (b) Eigenmode expansion simulation showing the optical power transfer from optical fiber to SiN waveguide. (c) Cross-sectional images of the mode shown in (b). (d) The dependence of the coupling efficiency on the SU8 cladding length for 2\,mm-long optical fiber taper.}
\end{figure}

Adiabatic coupling is an efficient technique to transfer an optical mode between nonuniform waveguide structures \cite{rasc1998}. Geometrical changes in the two waveguides are kept sufficiently gradual, so all optical power remains in the fundamental mode of the composite structure, and there is no energy transfer from the fundamental mode to higher-order modes. The lengths of the interacting waveguides should be much greater than the beat length $z_b=2\pi/(\beta_1-\beta_2)$, where \(\beta_1\) and \(\beta_2\) are the respective propagation constants \cite{Criteria}. In the present work, a terminating tapered fiber is prepared by first etching standard single-mode fiber into a taper and then cladding the tip in a higher-index SU8 polymer material. We refer to this cladding structure as a `cap'. In Sec. \ref{sec:Discussion} we also discuss terminating tapered fibers wherein the entire taper structure is covered in a higher-index polymer. We refer to such a structure as a `clad' tapered fiber. As light is propagating down the fiber, it is first guided by the index contrast between the core and cladding of the fiber. In the tapered region of the fiber, the polymer cap is gradually introduced before the fiber cladding is entirely etched away. In this region, light begins transferring from the fiber core to the polymer cap. Next, where the tapered fiber and polymer cap together have sufficiently small diameter that the mode extends beyond the polymer cap, care must be taken to ensure the evanescent tail does not couple to modes in the BOx. In this region, a small trench has been etched in the BOx leading up to the on-chip waveguide. Before the termination of the tapered fiber, the mode is entirely guided by the SU8 polymer cap with refractive index of 1.56 at 1550\,nm. Finally, the mode transitions from the polymer cap to the on-chip, high-index waveguide. In this work, silicon nitride (SiN) waveguides with index of 2.0 are studied on a BOx with index 1.46. A schematic of such a capped fiber meeting a waveguide is shown in Fig. \ref{fig:1stSchematic}(a).  

In the measurements under consideration, optical power is adiabatically transferred from the tapered fiber to the SU8 cap and then from the SU8 cap to the SiN waveguide, as shown in Fig.\,\ref{fig:1stSchematic}(a). The SU8 cap is necessary to ensure the effective index of the mode is higher than the index of the BOx in all regions of space wherein the evanescent field can reach the BOx.  If this condition is not satisfied the optical power will leak to the continuum of substrate modes. Design of this structure includes finding the optimal SU8 interaction length, and the initial transition to SU8 should not be abrupt. In simulation we have found leakage to substrate modes can be further reduced by etching a trench of 1.7\,\textmu m depth just before the beginning of the on-chip waveguide. This trench allows the evanescent tail of the supermode of the fiber/SU8 structure to extend beyond the material structure without coupling to undesired modes. The trench is etched around the initial segment of the waveguide to avoid an abrupt index transition (see Fig.\,\ref{fig:1stSchematic}(a)). 
 
The field distribution as a function of space is calculated by the eigenmode expansion (EME) method\cite{EME}. Figure \ref{fig:1stSchematic}(b) shows the EME simulation of 1550\,nm light propagating through the fiber-waveguide interface. In this simulation, the SiN waveguide was 220\,nm thick and tapering from 300\,nm to 1.5\,\textmu m in width over a length of 300\,\textmu m. The SU8 cap length used in this simulation was 100\,\textmu m, while the length of the tapered fiber was 2\,mm. The region in front of the SiN taper, where the optical mode was not protected by either the fiber cladding or the SU8 cap, was simulated with a trench etch depth of 1.7\,\textmu m. Light is launched in the fundamental mode of SMF-28e+ fiber. The light begins to transition to the SU8 while the fiber is over the etched trench, and the mode conversion is completed when the light transitions to the fundamental TE mode of the on-chip waveguide. Cross-sectional images of the mode at four locations along the structure are also shown in Fig. \ref{fig:1stSchematic}(c). The gradual, single-mode transfer of optical power from the fiber to the SiN waveguide is apparent. Figure \ref{fig:1stSchematic}(d) quantifies the dependence of the coupling efficiency on the SU8 cladding length.

\begin{figure}
\includegraphics[width=\linewidth]{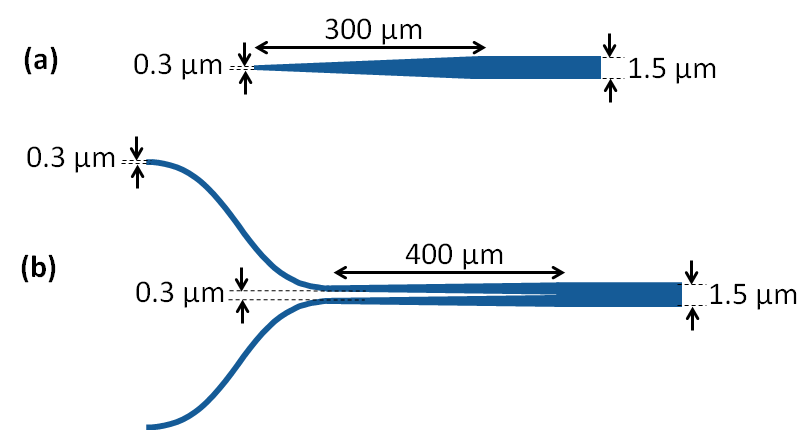}
\caption{\label{fig:Couplers} (a) Conventional tapered waveguide coupler. (b) Forked waveguide coupler.}
\end{figure}

In the simulations of Fig. \ref{fig:1stSchematic}, the on-chip waveguide begins with a taper. In practice, we have found coupling and packaging with such a taper to be sensitive to precise alignment. To alleviate alignment sensitivity, an alternative coupling structure based on a forked waveguide has also been considered. This forked coupler provides additional mechanical stability with a potential energy minimum at the center of the on-chip waveguide where the fiber is intended to reside. We have found empirically that the tapered fiber tends to sit symmetrically and firmly in the crook of the fork. This approach is less prone to movement under the influence of mechanical vibrations. The resulting package is more robust compared to the tapered coupler where the fiber is sitting on top of a sub-micron tapered waveguide. Furthermore, optical alignment is much easier, because the yaw and the pitch angles of the  fiber do not need to be as accurately tuned as in the case of the tapers. Figure \ref{fig:Couplers} shows the taper and fork waveguide couplers used in this report.

\section{\label{sec:Fabrication} Tapered Fiber Fabrication}
\begin{figure}
\includegraphics[width=\linewidth]{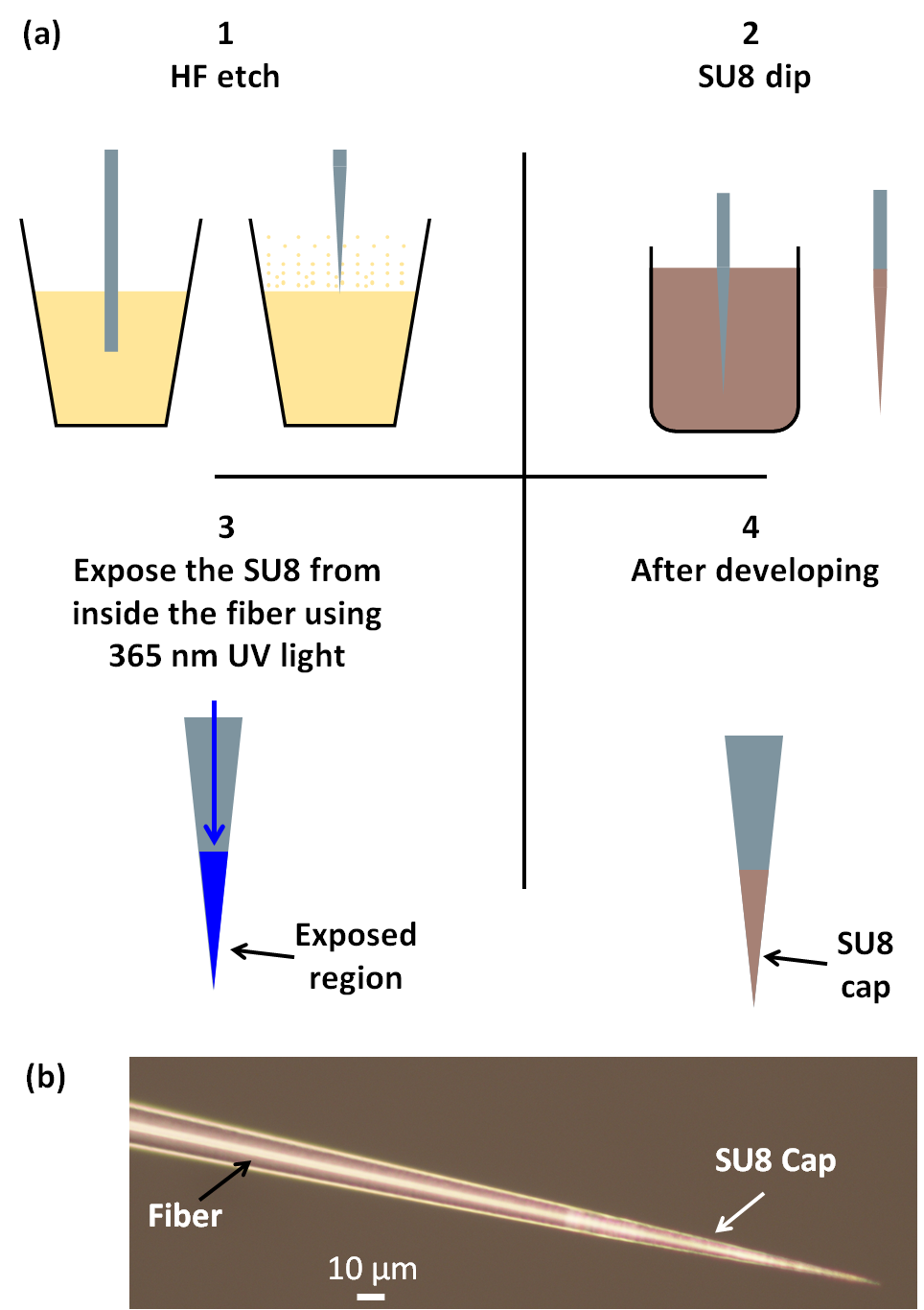}
\caption{\label{fig:Steps} a) Fabrication steps for SU8 capped tapered fiber. b) Microscope image of a SU8 capped tapered fiber}
\end{figure}

The terminating tapered fibers explored in this work were fabricated with a wet etching procedure using hydrofluoric acid (HF). Fabrication of tapered fibers with HF has previously been explored in the context of both continuous \cite{zhsa2010,kb2012,koja2019} and terminating tapers \cite{kb2012,mooh1996,saph1998,stfo1999,alst2000}. In this work we extend this fabrication to include a sub-wavelength polymer cladding to facilitate coupling to on-chip, high-index waveguides. Figure \ref{fig:Steps} shows the fabrication steps for the SU8 capped tapered fiber. To begin, the acrylic layer of the fiber was removed using hot sulfuric acid. This method of removal is necessary, as mechanical stripping leads to microcracks that are penetrate and deepen during HF etching. SMF-28e+ fiber was etched using a 35\,\% HF solution. The etch time was  2 hours (step 1 in Fig.\,\ref{fig:Steps}), and during this time both the liquid- and vapor-phase HF contribute to the profile of the etched fiber. The section of the fiber submerged in HF completely dissolves, while the section above the HF etches in a tapered profile with a sub-micron tip. The etching of the section of the fiber above the liquid is believed to result from a combination of liquid HF being drawn up the fiber by capillary action as well as etching by HF vapors. Both the liquid and vapor etchants decrease in concentration with height above the liquid, leading to a tapered fiber profile. A thorough clean with hot sulfuric acid was also necessary after the HF etch. The fiber taper was then inserted into SU8 photosensitive polymer (Step 2) and subsequently exposed using 365\,nm ultraviolet (UV) light. To produce a thin cladding on the surface of the fiber only around the tip (i.e., the cap), the SU8 polymer was exposed by sending UV light inside the fiber (step 3) using a fiber-coupled LED with a nominal wavelength of 365\,nm and output power of 9.8\,mW. However only a fraction of the LED power was available for the exposure due to the core diameter mismatch between the LED's multimode patch cord and the SMF-28e+ fiber. The core diameter of the multimode patch cord was 400\,\textmu m, and SMF-28e+ fiber was coupled to this multimode patch cord using a bare fiber adaptor. As the evanescent tail of the UV guided mode leaks outside the SiO$_2$ of the fiber taper, more SU8 polymer is gradually exposed down the length of the taper, leading to a polymer cap of gradually increasing thickness. In fabrication step 4, the fiber is developed, leaving an SU8 cap on the tip without an abrupt fiber/SU8 interface. A microscope image of a completed capped tapered fiber is shown in Fig.\,\ref{fig:Steps}(b).

\section{\label{sec:level1} Experimental Characterization}
\begin{figure}
\includegraphics[width=\linewidth]{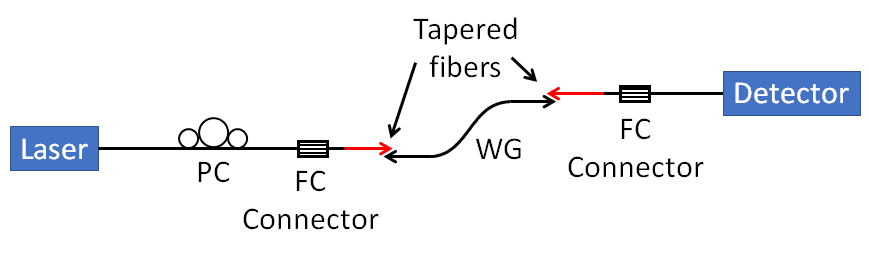}
\caption{\label{fig:Setup} Characterization setup. PC, FC, and WG stand for polarization controller, ferrule connector, and waveguide respectively.}
\end{figure}

The configuration employed for assessing coupling loss is shown in Fig. \ref{fig:Setup}. Light is coupled from a fiber, through one adiabatic taper, to an on-chip waveguide. Light then propagates through a section of the waveguide and off the chip through another adiabatic taper to a detector. The two fiber tapers are staggered to ensure that light is not passing directly from one fiber to the other without going through the waveguide. The length of the on-chip waveguide is 1.8\,mm. 

\begin{figure}
\includegraphics[width=\linewidth]{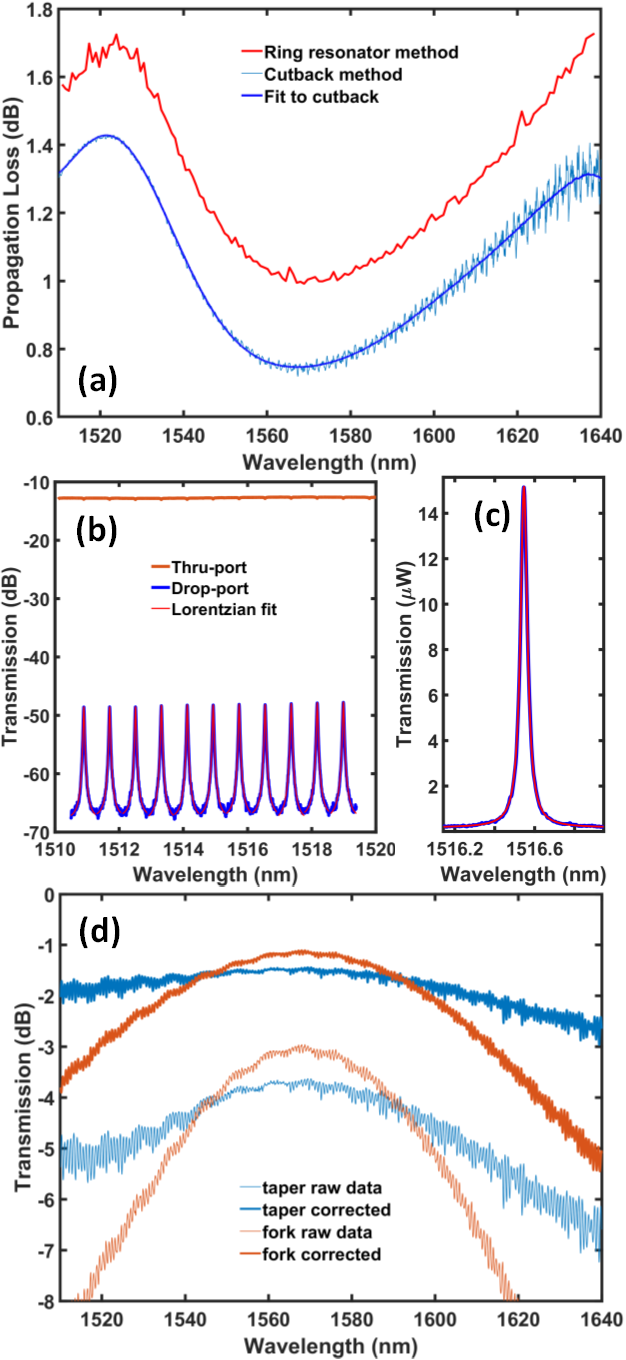}
\caption{\label{fig:MainPlot} (a) Propagation losses in a 1.8\,millimeter-long SiN waveguide over wavelengths of interest. The light blue line is the loss measured by cutback method and the dark blue line is a polynomial fit used for the normalization of the transmission spectrum to extract the coupling loss. The red line is the propagation loss estimated from the \textit{Q} of an under-coupled ring resonator. (b) Transmission through drop- and thru-port of the ring resonator, showing that the extinction ratio is more than 30\,dB. (c) Lorentzian curve fitting to the experimental data for estimation of the \textit{Q} factor. (d) Transmission spectra. The thin, light blue (red) line is the spectrum of the taper (fork) device. The thick blue (red) line is the spectrum of a single coupler extracted by correcting to account for propagation loss and dividing by two to obtain the loss per coupler.}
\end{figure}

To extract coupling loss, the propagation loss of this waveguide section needs to be measured accurately. We measured the propagation loss with two independent techniques. One technique extracts the propagation loss from the \textit{Q} factor of an under-coupled ring resonator, using $\alpha\approx2\pi n_g/Q\lambda_0$, where \(\alpha\), \(n_g\), and \(\lambda_0\) is the loss per unit length, group index, and wavelength, respectively\cite{payam}. The other technique measures loss of progressively longer waveguides, called the cutback method. The waveguide lengths used for the cutback method varied from 2\,mm to 21.8\,mm, in steps of 3.3\,mm. Figure \ref{fig:MainPlot} shows waveguide propagation loss and coupling insertion loss. The red trace in Fig.\,\ref{fig:MainPlot}(a) shows the propagation loss in a 1.8\,millimeter-long SiN waveguide as measured by the \textit{Q} of the ring resonator. The radius of the ring resonator used was 250\,\textmu m, and all SiN structures in this study were 220\,nm thick. The SiN was deposited using plasma-enhanced chemical-vapor deposition, as described in Ref.\,\onlinecite{shbu2017}. The ring resonator was under-coupled with an extinction ratio of more than 30 dB, as shown in Fig.\,\ref{fig:MainPlot}(b). Figure\,\ref{fig:MainPlot}(c) shows the Lorentzian curve fitted to the experimental data. The blue trace in Fig.\,5(a) is the propagation loss obtained with the cutback method. From the plot we see the cutback method gives a significantly lower value of propagation loss. The cutback method data were used to estimate the coupling loss to ensure a conservative estimate for fiber-to-waveguide insertion loss. The reason for the discrepancy between the \textit{Q} and cutback methods is not known.

Before measuring the transmission, the experimental setup was normalized for wavelengths from 1460\,nm to 1510\,nm by replacing the device under test with a fiber patch cord between the two FC connectors, as shown in Fig. \ref{fig:Setup}. Figure \ref{fig:MainPlot}(d) shows the transmission spectra through taper and fork waveguide devices. The minimum measured values of loss per coupler after correcting for waveguide propagation loss in the optical path are 1.1\,dB per coupler for the fork coupler and 1.4\,dB per coupler for the taper. The 3\,dB-bandwidth of the fork coupler is 90\,nm, while the taper coupler has a 1\,dB-bandwidth of more than 100\,nm. The extrapolated 3\,dB-bandwidth is more than 250\,nm.

\begin{figure}
\includegraphics[width=\linewidth]{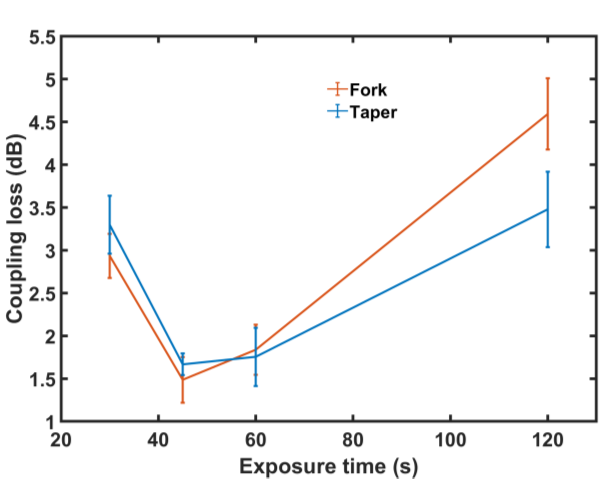}
\caption{\label{fig:Variations} Coupling loss versus SU8 UV exposure time for taper and fork couplers.}
\end{figure}

The coupler insertion loss depends on the profile of the SU8 cap. This profile can be adjusted  by the UV exposure time, with longer exposure leading to a thicker, longer cap. Figure \ref{fig:Variations} shows the loss through fork and taper couplers as a function of the exposure time.  The error bars in this plot were obtained by manufacturing multiple tapers in different fabrication batches and measuring their performance when coupling to the same waveguide structure. The error bars represent the standard deviation of these measurements. Variation in coupler loss was due to wet etching and manual application of the SU8 on the fiber by dipping. The amount of SU8 remaining on the fiber taper before exposure and development depends on the speed at which the fiber is removed from the SU8. Fast removal results in more SU8 remaining on the fiber. This excess gathers into a bead at the fiber tip. This bead is difficult to remove with the standard developing procedure and leads to decreased coupling efficiency. One important insight from Fig.\,\ref{fig:Variations} is that a relatively broad range of exposure times provides similar coupling, indicating robustness with respect to this processing condition.

\section{\label{sec:Packaging} Packaging}
For many applications, coupling light from a fiber to a waveguide is only useful if the assembly can be packaged in a manner that enables deployable systems. This is often accomplished by aligning fibers to coupling structures and fixing them in place with epoxy. For the tapered fibers under consideration, we do not use epoxy in the evanescent coupling region because, to our knowledge, all epoxies have an index of refraction very close to that of the SU8 used as a cap and would lead to high losses. We have therefore used an approach to packaging wherein each tapered fiber is fixed with epoxy in a groove etched in a separate silicon chip. The tapered region of the fiber extends beyond the carrier chip so that it can be placed on the waveguide device chip.

\begin{figure}
\includegraphics[width=\linewidth]{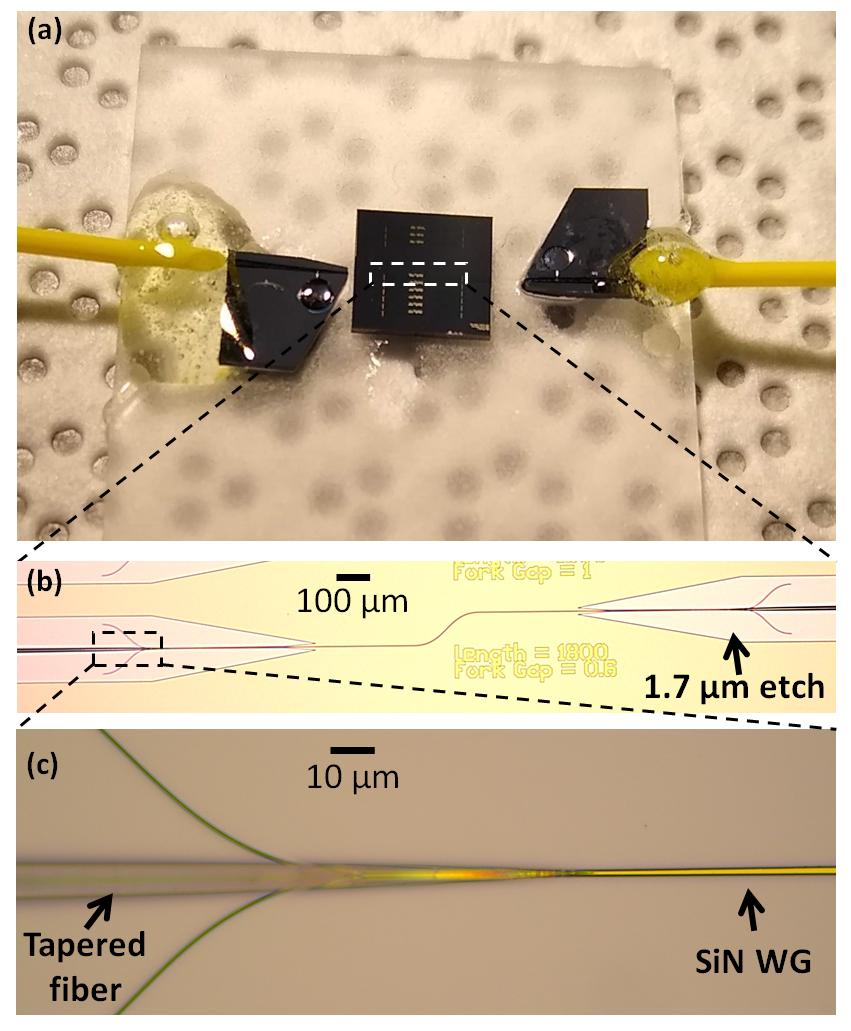}
\caption{\label{fig:Package} Images of the packaged device. (a) A packaged device on a glass slide. Fibers on carrier chips are seen on either side. (b) An $s$-shaped 1.8\, millimeter-long SiN waveguide with fork couplers and tapered fiber on either side. (c) SU8 capped tapered fiber sitting on the fork waveguide coupler in the packaged device.}
\end{figure}

\begin{figure}
\includegraphics[width=\linewidth]{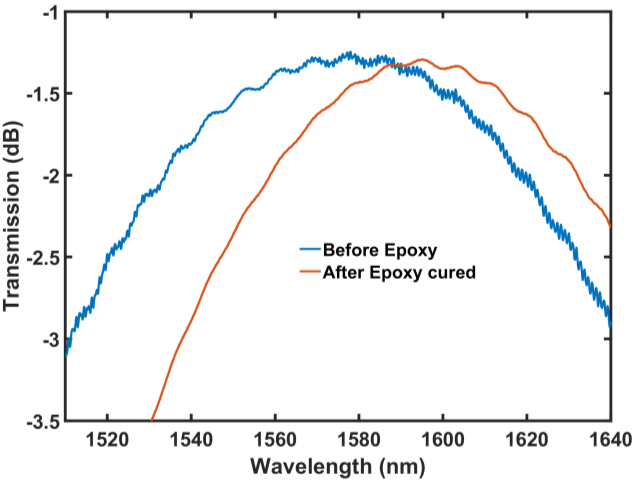}
\caption{\label{fig:TransAfterPackage} Transmission spectrum of the fork coupler before the application of epoxy underneath the carrier chip (blue line) and after the epoxy has been cured (red line).}
\end{figure}

The silicon carrier chips were fabricated by etching channels 50\,\textmu m deep and 50\,\textmu m wide using deep reactive ion etching. The tapered fibers were epoxied to the carrier silicon chips using Stycast 1266. After placing the fiber into the channel, epoxy was delivered to the channel through a delivery compartment etched on the same fiber carrier chip. For packaging, the device chip was first mounted on a carrier substrate, again with Stycast 1266. A glass slide was used for this purpose, but other substrates may suffice as well. When aligning the tapered fiber to the device chip, the carrier chip was held by vacuum suction. After the tapered fiber was aligned to the device chip, the carrier chip was fixed in place on the carrier substrate. To perform this step, Stycast 1266 was applied on the glass slide close to the carrier chip. Due to surface tension, the epoxy is drawn between the carrier chip and the glass slide and spreads uniformly under the carrier chip. Height adjustment to realign the fiber is often needed after the application of the epoxy. The alignment and packaging procedure are conducive to automation with machine vision and feedback from optical transmission. Figure \ref{fig:Package}(a) shows a packaged device, while Fig.\,\ref{fig:Package}(b) shows a zoom of the $s$-shaped 1.8\,mm-long SiN waveguide with fork couplers and tapered fibers on either side. Figure \ref{fig:Package}(c) shows the SU8 capped tapered fiber aligned on the fork waveguide coupler in the packaged device. This devices was measured before and after epoxy was applied and cured. Figure \ref{fig:TransAfterPackage} shows the transmission spectrum of the fork coupler before the application of epoxy underneath the carrier chip (blue line) and after the epoxy has been cured on both sides (red line). The spectrum shifts due to a slight adjustment in the longitudinal position of the tapered fiber relative to the fork coupler. This adjustment can be pre-compensated prior to delivering and curing the epoxy.

\section{\label{sec:Discussion}Summary and Discussion}
This work has explored capped, terminating, tapered fibers for optical coupling from a fiber to an on-chip waveguide. The tapered fibers have been capped with higher-index polymer to enable coupling to waveguides located on a buried oxide substrate without leakage to substrate modes. The fibers have been used to couple to both tapered and forked on-chip waveguide structures. Low insertion loss has been achieved across a broad bandwidth. The approach to fiber-to-chip coupling has been used to package a device, indicating the potential for use in  research as well as industrial contexts with minimal additional insertion loss.

Several opportunities exist for follow on work. To further streamline the fabrication process in industrial settings, it may be advantageous to eliminate the exposure of the SU8 polymer all together. This may be possible if the entire fiber tip is cladded in SU8 of a suitable thickness. For this approach to achieve high optical transmission, the cladding thickness must be accurately tuned. The cladding must be thin enough that the effective index of a mode in a polymer slab of that thickness is lower than the effective index of the fiber mode. However, the cladding must be thick enough that when the polymer surrounding the fiber converges at the fiber tip, the effective index of this thicker polymer structure is higher than the effective index of the fiber mode. Achieving this thickness appears possible through appropriate dilution of the SU8 prior to the fiber dip. Simulations suggest such an approach is also promising, but we have not thoroughly explored this technique.

Using a polymer cladding may not be ideal for this fiber coupling approach, as the index of polymers tends to be very close to the index of epoxies used in packaging. It may be more advantageous to use high-index dielectrics such as amorphous silicon or silicon nitride to form the high-index cap. It may be possible to deposit such materials on the etched fiber tapers, thereby enabling the use of epoxy of index near 1.56 to be applied directly to the tip of the fiber without adding optical loss. Such a use of epoxy is likely to greatly improve the success of packages during cryogenic cycling, as is required in applications that seek to utilize superconducting single-photon detectors integrated with on-chip photonic components. The packages presented here have been tested at cryogenic temperature of 4\,K. On most trials, lateral misalignment of the fiber caused the insertion loss to increase by $\approx$\,5\,dB compared to room temperature. This lateral movement of the fiber during cooling could be eliminated if epoxy could be delivered right at the point where the fiber meets the waveguide. Dielectric claddings may also be more suitable for high-optical-power applications, as dielectrics will have higher damage thresholds than polymers.



\section*{\label{sec:Ack} Acknowledgements}

This work was funded by the IARPA SuperCables program. This is a contribution of NIST, an agency of the US government, not subject to copyright.

\section*{\label{sec:References} References}
\bibliography{Ref}
\end{document}